\documentclass[aps,prb,twocolumn,superscriptaddress,amsmath,amssymb]{revtex4}
\usepackage{graphicx} 
\usepackage{epsfig} 
\usepackage{float}
\usepackage{dcolumn}
\usepackage{bm}

\begin{document}

\hsize=6.5 true in
\vsize=8.9 true in
\parindent =15 pt
\hoffset= 0.1 in

\title{ Modifications and Extensions to Harrison's Tight-Binding Theory}
                                                                                
\author{  Lei Shi } 

\affiliation{ School of Computational Sciences, George Mason University, Fairfax, VA 22030}

\author{Dimitrios A. Papaconstantopoulos }

\affiliation{ School of Computational Sciences, George Mason University, Fairfax, VA 22030}
\affiliation{ Center for Computational Materials Science, Naval Research Laboratory, Washington, DC 20375}

\begin{abstract}
Harrison's tight-binding theory provides an excellent qualitative description
of the electronic structure of the elements across the periodic table. However,
the resulting band structures are in significant disagreement with those 
found by standard methods,particularly for the transition metals. For these 
systems we developed a new procedure to generate both the prefactors
of Harrison's hopping parameters and the onsite energies. Our approach gives an
impressive improvement and puts Harrison's theory on a quantitative basis. Our
method retains the most attractive aspect of the theory, in using a revised set
of universal prefactors for the hopping integrals. In addition, a new form of
onsite parameters allows us to describe the lattice constant dependence of the
bands and the total energy, predicting the correct ground state for all
transition, alkaline earth and noble metals. This work represents not 
only a useful computational tool but also an important pedagogical 
enhancement for Harrison's books.
\end{abstract}

\maketitle

\section{Introduction}
Walter Harrison developed an elegant analytic theory of the electronic 
structure of solids \cite{har_book_80,har_book_99} . This 
theory has been very sucessful in providing a physical 
understanding of the electronic structure and the characteristics of bonding. 
However, Harrison's theory of solid state has limited ability to 
produce accurate numerical results for the band structure, density of states 
and the relative stability of different crystal structures.

\smallskip

In this work, we have set out to put Harrison's approach on a quantitative
foundation. We have now realized that it is possible to put the Slater-Koster
parameters in the form given by Harrison but with new prefactors and 
determine new onsite parameters. The result is that we retain the 
universality of Harrison's parameters, which means  the same prefactors 
for all transition,alkaline earth and noble metals, but with different
onsite terms for each element. It is clear to us that this approach,perhaps
slightly modified,may be extended to cover the rest of the periodic table. 
We have also succeeded with a small number of additional
parameters to describe the volume and structure dependence of the energy
bands and, therefore, obtain total energies and predictions of relative 
stability.

\smallskip
Harrison has opted for simplicity in the LCAO approach and has created a
set of universal hopping parameters that can easily be used to perform
calculations. In the tables of his books,Harrison uses atomic energies 
as onsite parameters in his Hamiltonians, which is the main shortcoming 
of directly using these tables, to perform sufficiently accurate band 
structure calculations. However,Harrison pointed out(\cite{har_book_99},p561) 
that atomic values for the d-state energies need to be corrected for 
differences in d-state occupancy, and gave a way for doing that in the 
case of Cr.
\smallskip
                                                                                
\begin{figure}[ht]
\begin{minipage}[c]{.24\textwidth}
\centering
\includegraphics[width=1.5in,height=1.7in]{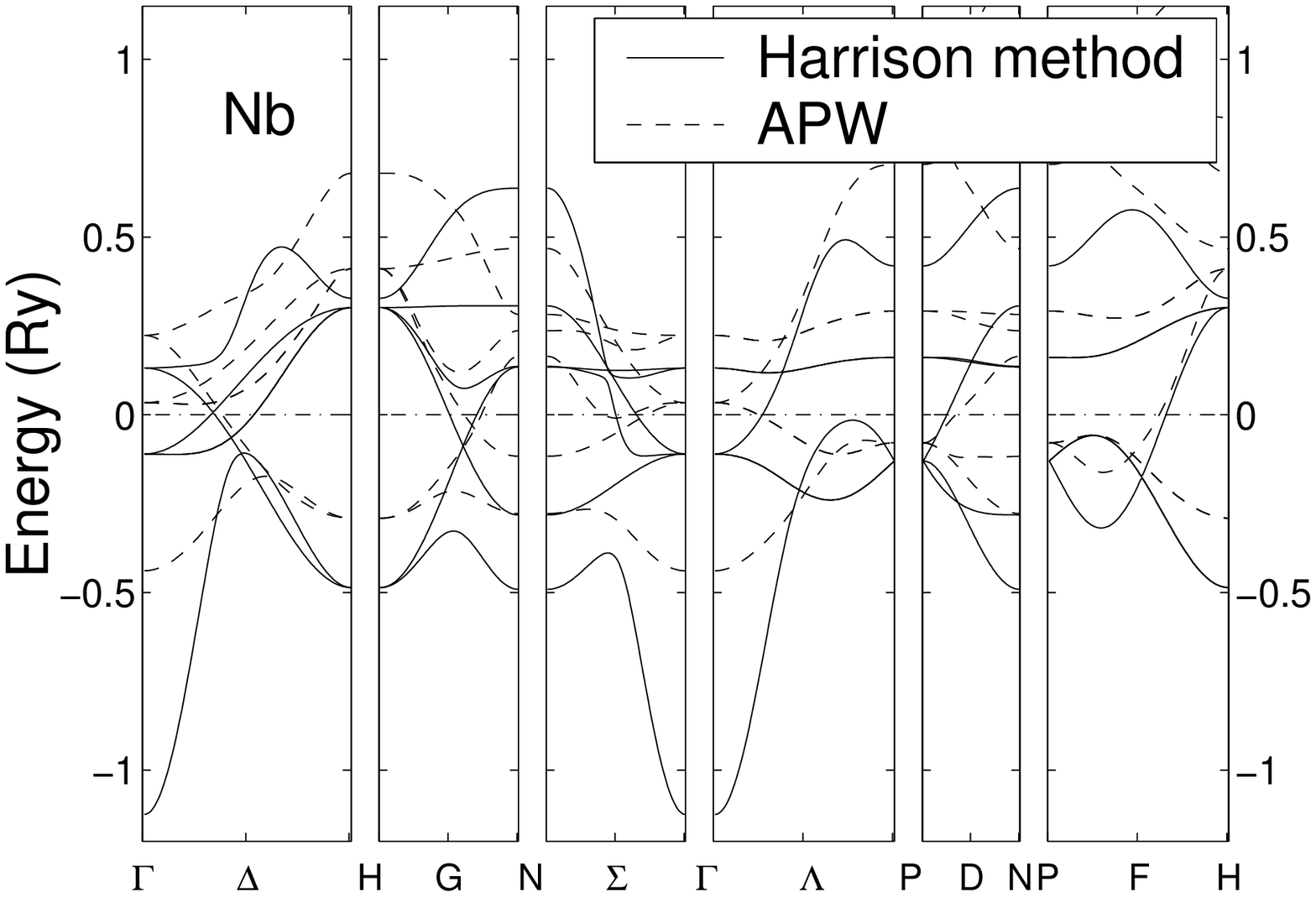}
\end{minipage}%
\begin{minipage}[c]{.24\textwidth}
\centering
\includegraphics[width=1.5in,height=1.8in]{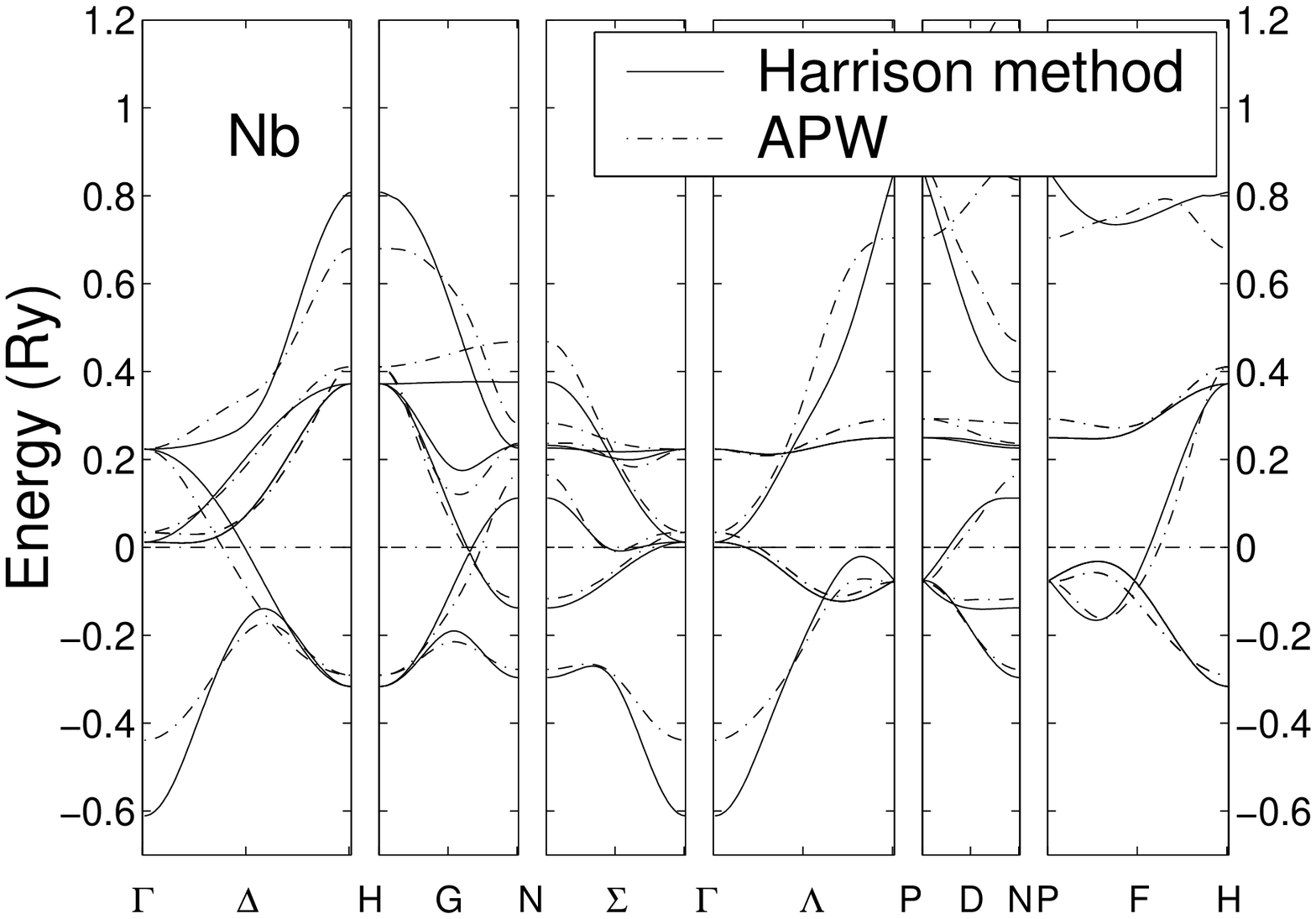}
\end{minipage}%
\caption{Harrison's energy bands of Nb compared to APW. Left graph created by \( 6\times6 \) 
Hamiltonian; right graph created by \( 9\times9 \) Hamiltonian and fitted onsite energies. } 
\label{har_band_fit}
\end{figure}

We illustrate the importance of correcting the values of the onsite parameters
for the transition metals Nb and Pd by using Harrison's hopping parameters and
uncorrected atomic term values. We compared the 
results of a \( 6\times6 \) Harrison Hamiltonian(without \( p \) orbitals) as given
in Harrison's book and we found that the energy bands created this way are in 
serious disagreement with Augmented Plane Wave(APW) results
(see left Fig.~\ref{har_band_fit}). We also tested a \( 9\times9 \)(with \( p \) orbitals) Harrison Hamiltonian with all 
hopping prefactors kept at 
Harrison's values, but the onsite parameters modified by fitting the 
energy bands to APW calculations. \cite{Sigalas,papa} This modification 
gave us 
better 
results in the \( d- \)bands, 
but there was still a large error for the s-like first band(see right 
Fig.~\ref{har_band_fit}). 
Our conclusion is that Harrison's theory 
can only give a qualitative description of the band structure of the 
transition metals even if we fit the onsite terms to first-principles 
results.  

\smallskip
\section{Energy Bands and Density of States}
\begin{figure}[htp]
\begin{minipage}[c]{.26\textwidth}
\centering
\includegraphics[width=1.7in,height=2.0in]{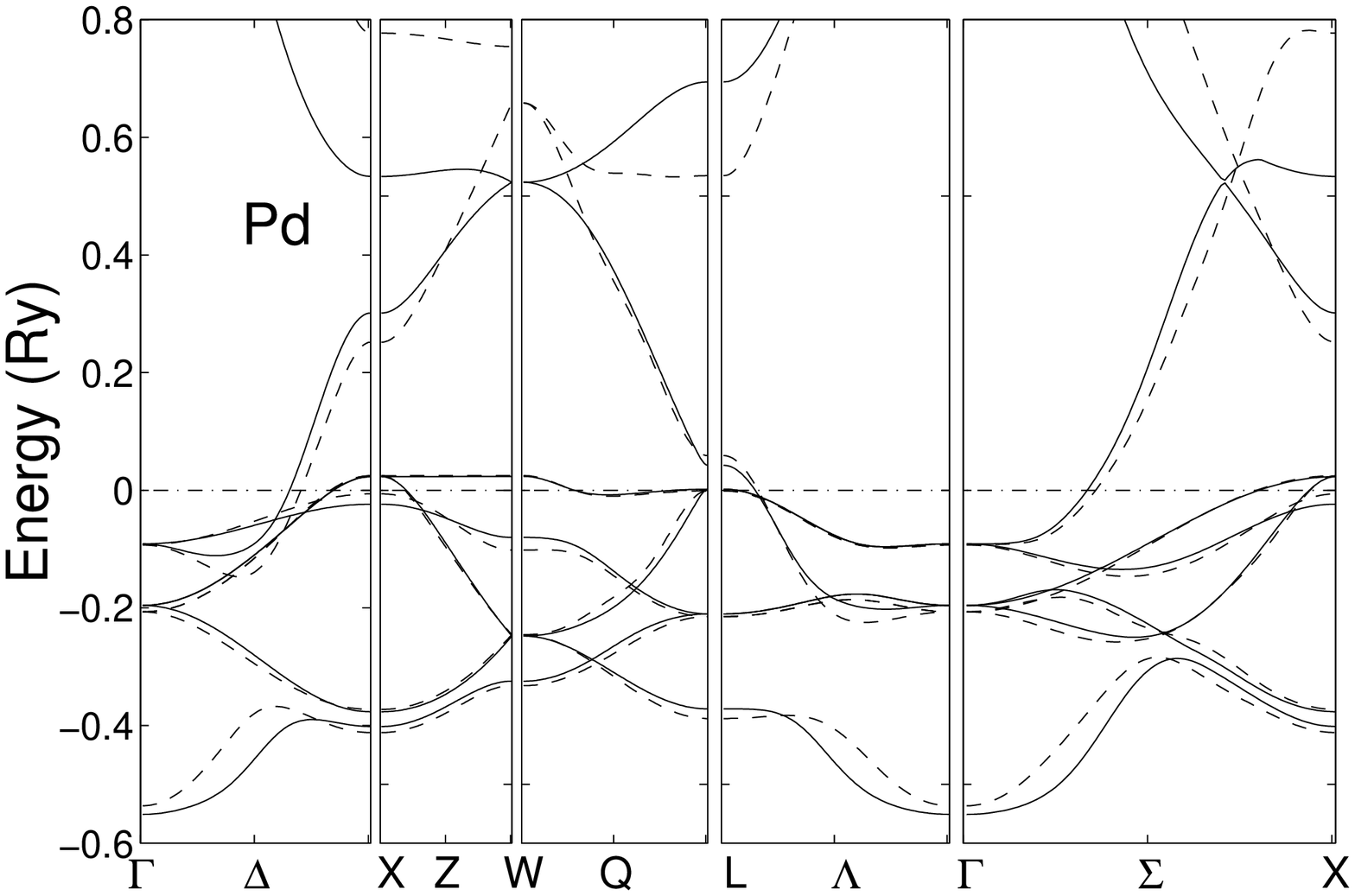}
\end{minipage}%
\begin{minipage}[c]{.26\textwidth}
\centering
\includegraphics[width=1.7in,height=2.0in]{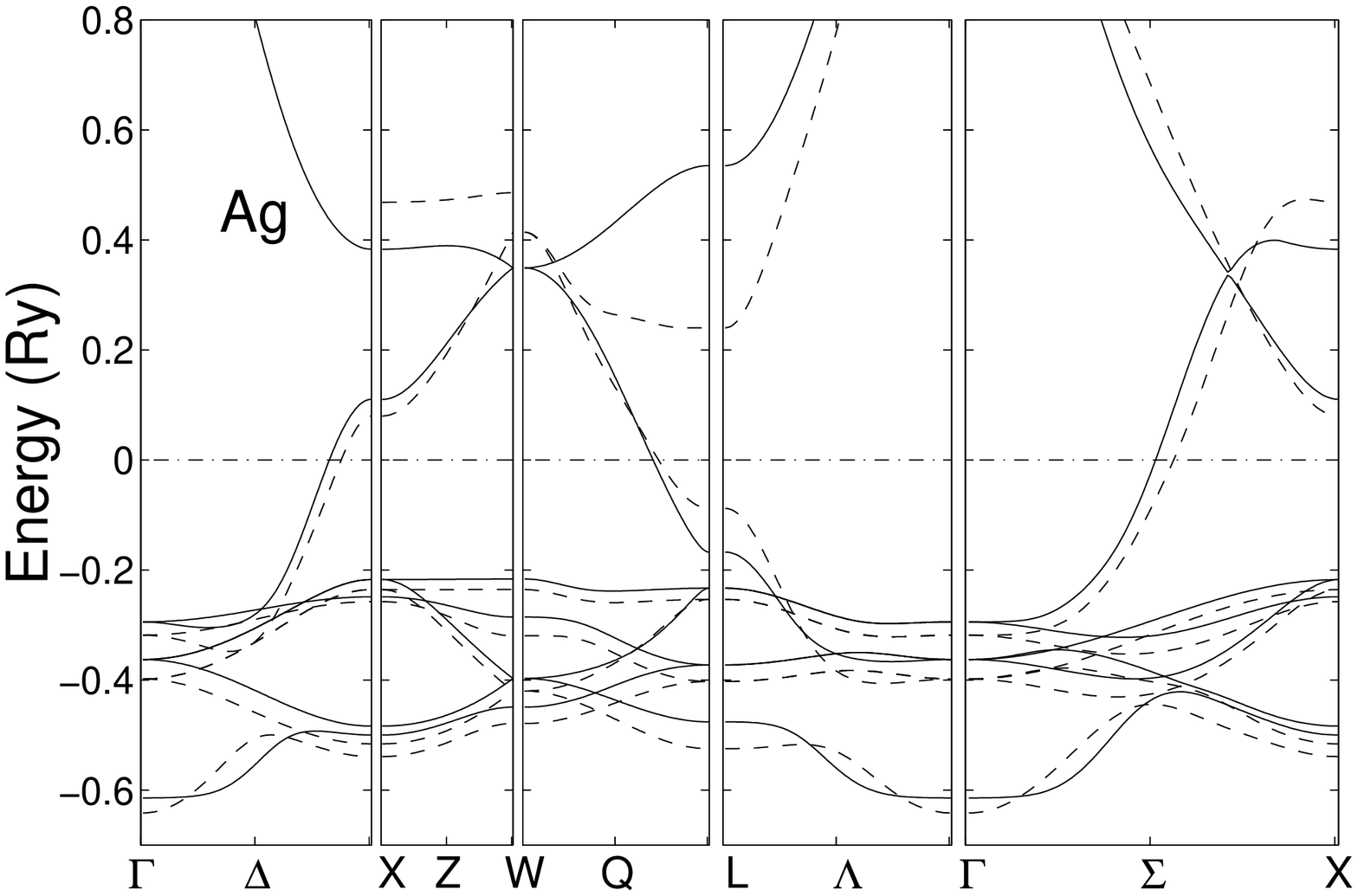}
\end{minipage}%
                                                                                
\smallskip
                                                                                
\begin{minipage}[c]{.26\textwidth}
\centering
\includegraphics[width=1.7in,height=2.0in]{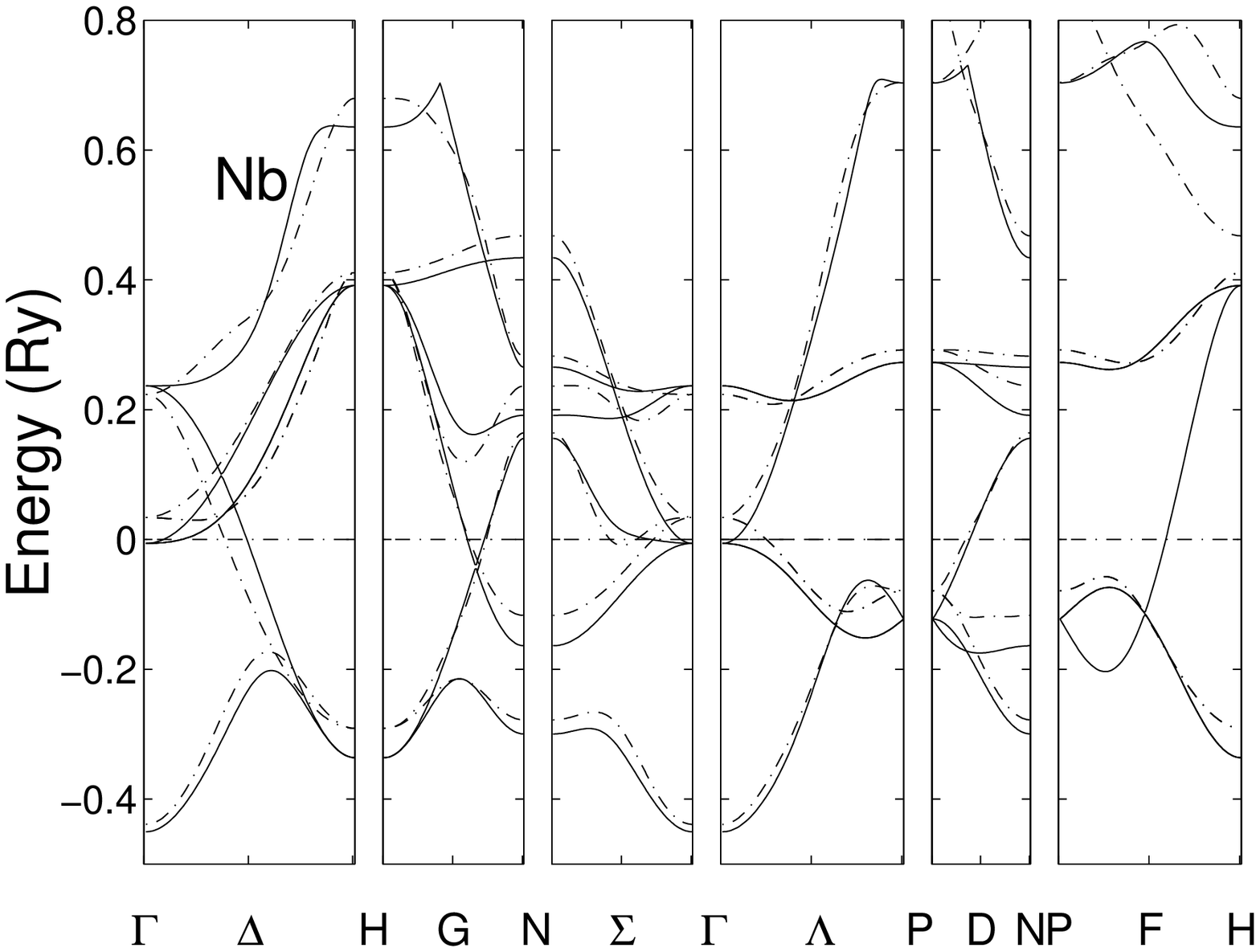}
\end{minipage}%
\begin{minipage}[c]{.26\textwidth}
\centering
\includegraphics[width=1.7in,height=1.9in]{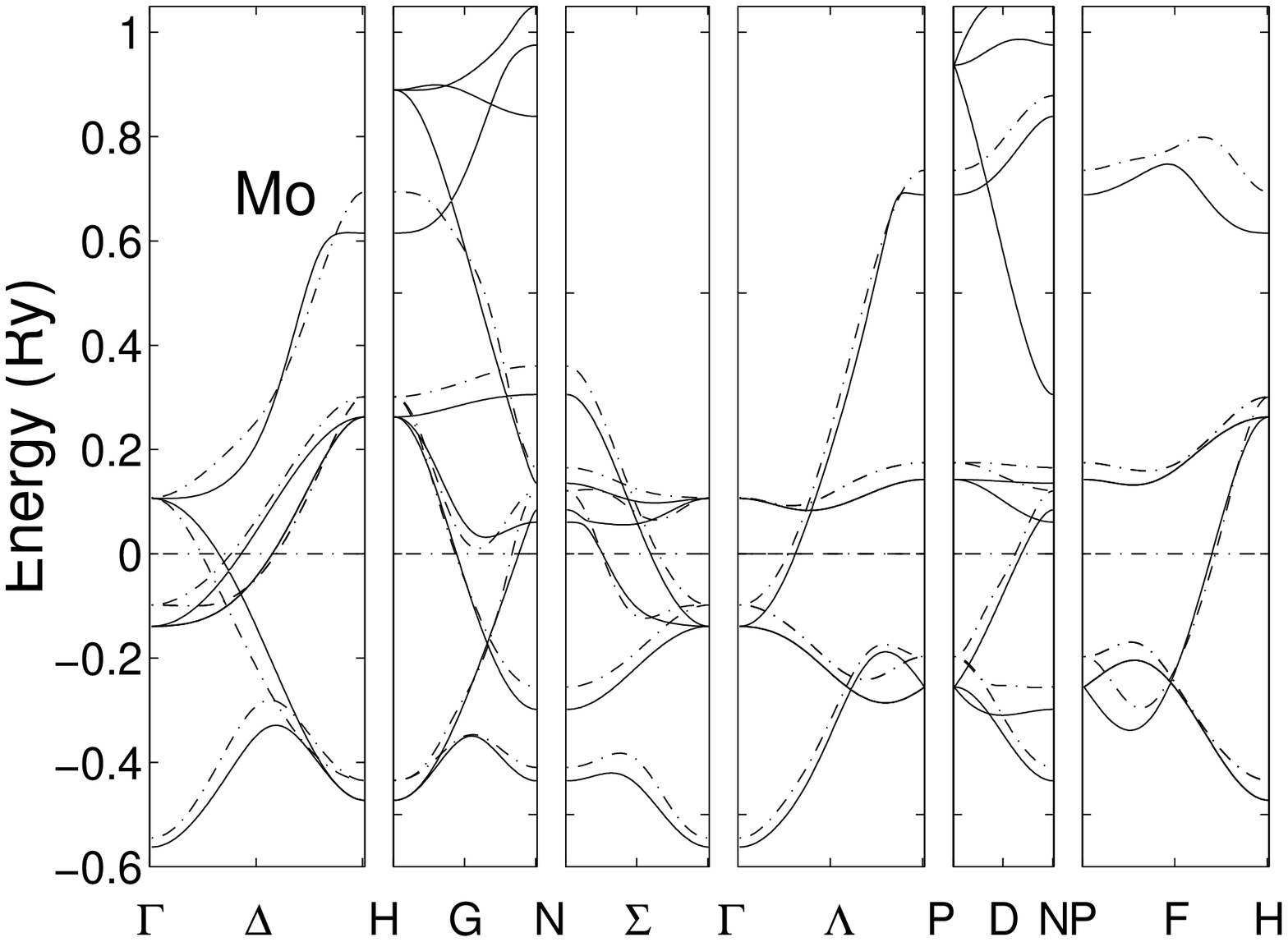}
\end{minipage}
\caption{APW and Modified Harrison's energy bands of Ag, Pd, Nb and Mo. The 
solid line is the modified Harrison result and dash-dotted line is the 
APW result.}\label{bands_we}

\bigskip

\begin{minipage}[c]{.26\textwidth}
\centering
\includegraphics[width=1.7in,height=2.0in]{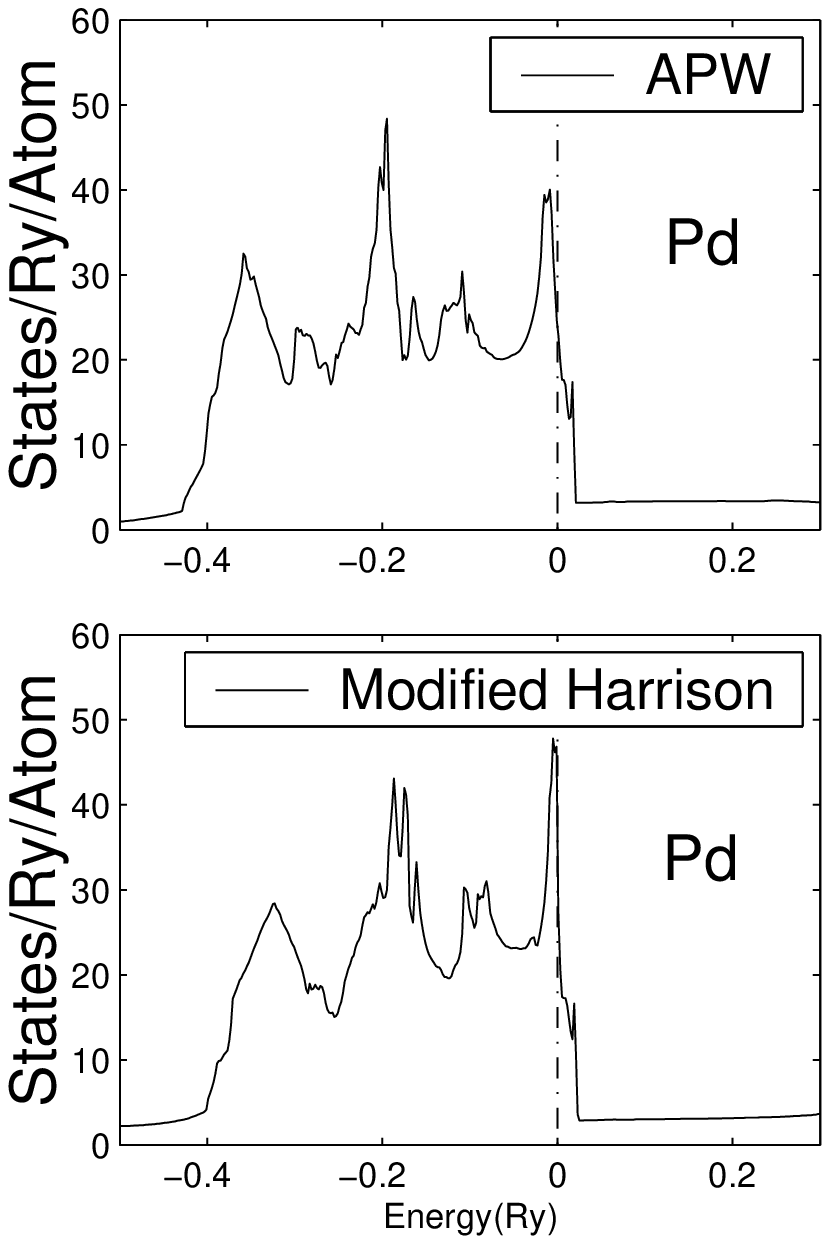}
\end{minipage}%
\begin{minipage}[c]{.26\textwidth}
\centering
\includegraphics[width=1.7in,height=2.0in]{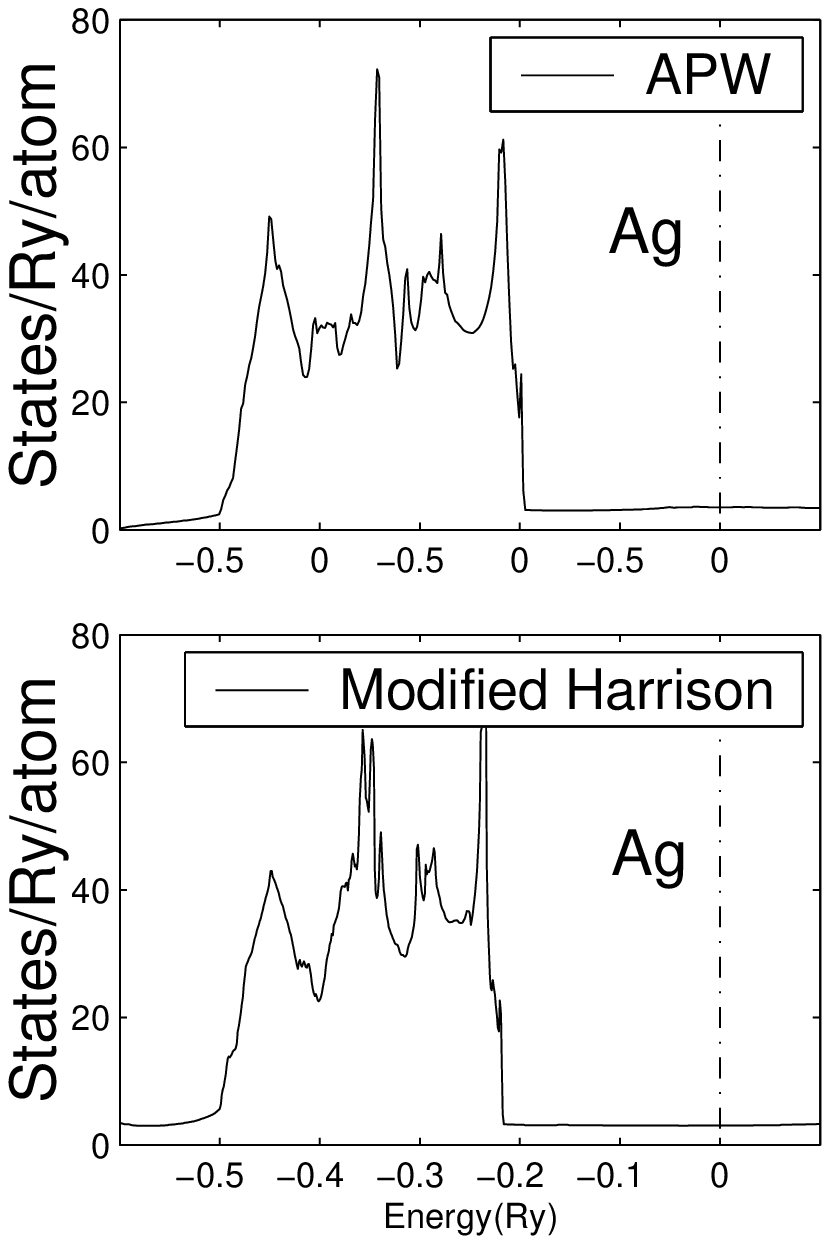}
\end{minipage}%

\smallskip

\begin{minipage}[c]{.26\textwidth}
\centering
\includegraphics[width=1.7in,height=2.0in]{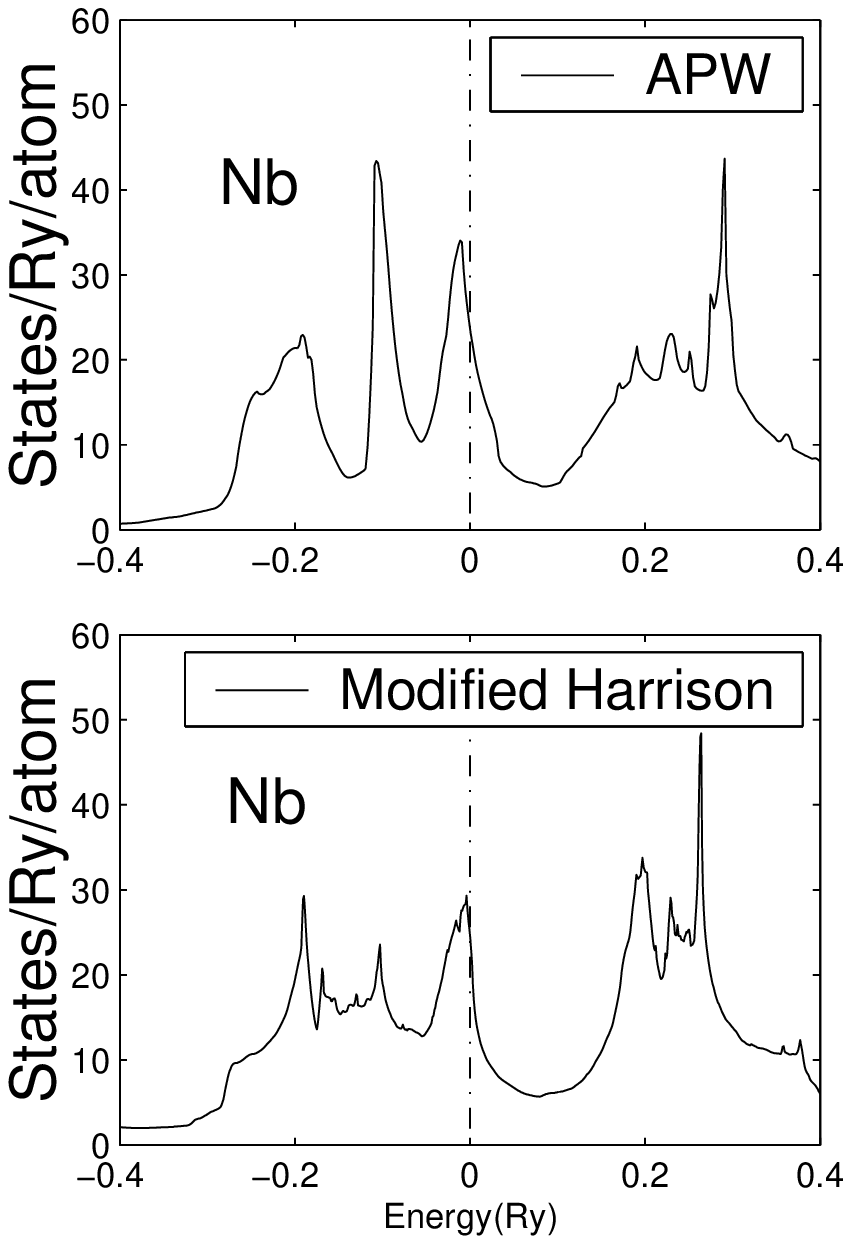}
\end{minipage}%
\begin{minipage}[c]{.26\textwidth}
\centering
\includegraphics[width=1.7in,height=2.0in]{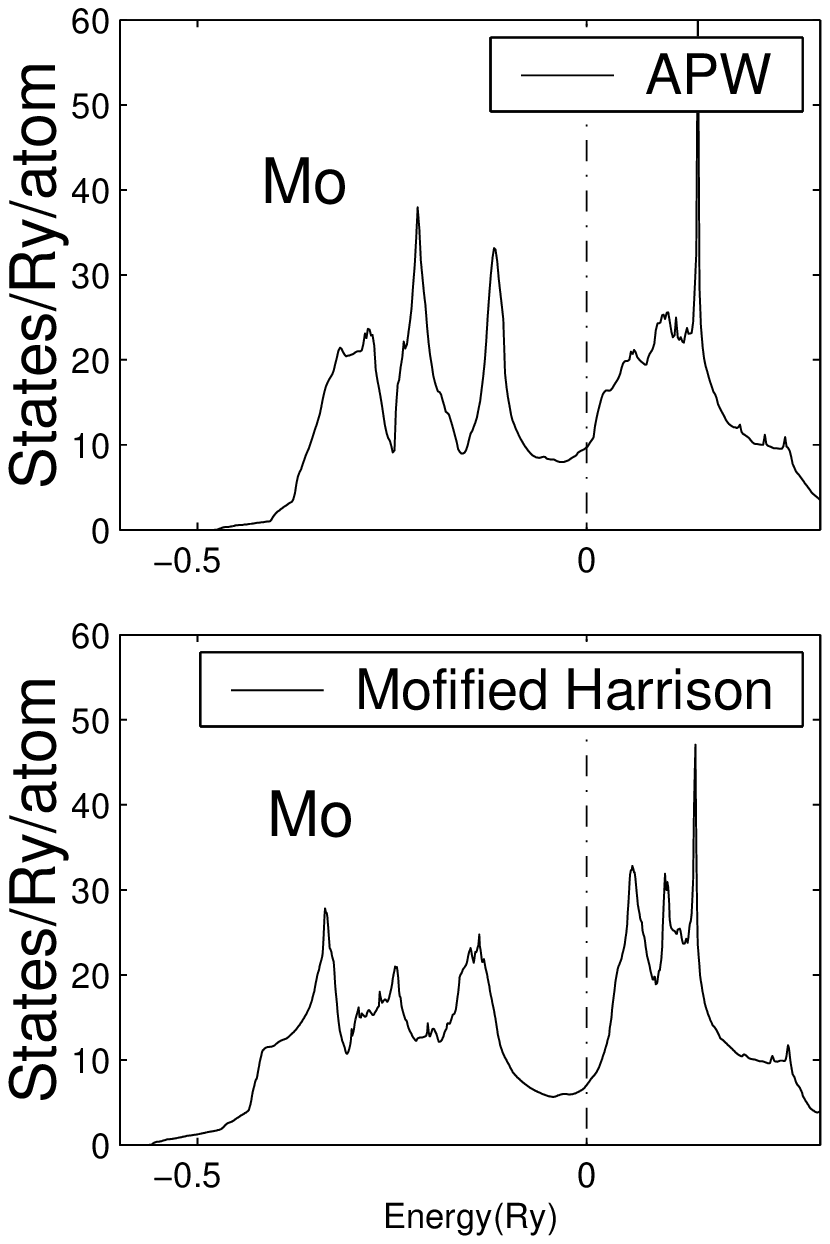}
\end{minipage}
\caption{APW and Modified Harrison's density of states of Ag, Pd, Nb and Mo.}\label{dos_we}

\end{figure}

We have developed a procedure that while maintaining the simplicity 
of Harrison's 
approach gives an impressive improvement that puts the theory on a 
quantitative basis. To accomplish this we have made the following 
modifications to Harrison's theory: (1) We introduced a \( p \) onsite 
energy as an additional parameter to the 
\( s \) and \( d \) onsite energies used by Harrison, and fit them all 
to APW results. (2) 
We modified the \( sp \) hopping integrals of Harrison, by introducing
 a dimensionless parameter 
\( \gamma_{s} \) as follows:
\begin{equation}
V_{ll'm}=\eta_{ll'm}\frac{\gamma_{s}\hbar^{2}}{m_{e} d^{2}}
\end{equation}
The parameter \( \gamma_{s} \) provides more flexibility to fit the first 
and sixth bands. (3)
We obtained new hopping prefactors by simultaneously fitting
the APW energy bands of the following 12 transition metals: V, Cr, Ni, Cu,
Nb, Mo, Pd, Ag, Ta, W, Pt and Au. In this fit, all 12 elements have the same 
common prefactors \( \eta_{ll^{\prime}m} \) but different(for each element)
onsite energies s, p and d, and also different values for the parameters 
\( \gamma_{s} \) and \( r_{d} \) that appear in the hopping parameters. 
Our Hamiltonian corresponds 
to an orthognal basis set as in Harrison. We did the above fitting at the
equilibrium lattice constants of the structure, which is the ground
state of each element, and included interactions of nearest, second-nearest,
and third-nearest neighbors for the
\( bcc \) structure and nearest and second-nearest for the \( fcc \) structure.
Of course, using more neighbors than Harrison did in the fit would automatically
make changes in the parameters, even if one were fitting the same bands.
Using the parameters determined with the above procedure, we reproduced APW  
energy bands and density of states(DOS) remarkably well, not only for the 
12 elements originally fitted, but also for the rest of the 
transition metals, the alkaline earth and the noble metals, as seen in 
Fig.~\ref{bands_we} and Fig.~\ref{dos_we} for four of the elements.  
Our new Hamiltonian prefactors, common for all metals,
together with Harrison's original prefactors are shown in 
Table~\ref{we_sst}. The onsite terms and the parameters 
\( \gamma_{s} \) and \( r_{d} \) for each element are shown in Table~\ref{onsite}. 

\smallskip


\begin{table}[ht]
\caption{ Harrison's hopping prefactors and our modified values.}\label{we_sst}
\begin{ruledtabular}
\begin{tabular}{|c|c|c|c|c|c|} 
   & \( \eta_{ss\sigma} \) & \( \eta_{sp\sigma} \) & \( \eta_{pp\sigma} \) & \( \eta_{pp\pi}    \) & \( \eta_{sd\sigma} \) \\ \hline
Harrison & -1.32 & 1.42 & 2.22 & -0.63 & -3.16 \\
Modified Harrison & -0.90 & 1.44 & 2.19 & -0.03 & -3.12 \\ \hline
& \( \eta_{pd\sigma} \) & \( \eta_{pd\pi}    \) & \( \eta_{dd\sigma} \) & \( \eta_{dd\pi}      \) & \( \eta_{dd\delta} \) \\ \hline
Harrison & -2.95 & 1.36 & -16.2 & 8.75 & -2.39 \\
Modified Harrison & -4.26 & 2.08 & -21.22
&  12.60 & -2.29 \\  
\end{tabular}
\end{ruledtabular}
\end{table}


\begin{table}[ht]
\caption{ Onsite parameters, \( \gamma_{s} \) and \( r_{d} \).}\label{onsite}
\begin{ruledtabular}
\begin{tabular}{|c|c|c|c|c|c|} 
Name  & \( s(Ry) \)  & \( p(Ry) \)  & \( d(Ry) \)   & \( \gamma_{s} \) & \( r_{d}(a.u.) \) \\ \hline
K        & 0.26067  & 0.22200  & 0.25426   & 1.13081 & 3.65981    \\
Ca       & 0.11994  & 0.24522  & 0.03657   & 1.07535 & 2.66618    \\
Sc       & 0.14809  & 0.38903  & -0.06695  & 0.98860 & 2.12358    \\
Ti       & 0.51352  & 0.79759  & 0.21879   & 0.92307 & 1.85267    \\
V        & 0.64331  & 0.73136  & 0.04711   & 0.90164 & 1.65358    \\
Cr       & 0.76372  & 0.86088  & 0.06389   & 0.87733 & 1.51087    \\
Mn       & 0.47377  & 0.86874  & -0.00300  & 0.82491 & 1.42366    \\
Cu       & 0.54432  & 0.93013  & -0.05425  & 0.92178 & 1.23548    \\
Zn       & 0.44779  & 0.77968  & -0.10598  & 0.78430 & 0.97054    \\
         &          &          &           &         &            \\
Sr       & 0.32339  & 0.41296  & 0.22810   & 1.22463 & 3.29024    \\
Y        & 0.29652  & 0.51778  & -0.07450  & 1.19367 & 2.75073    \\
Zr       & 0.54322  & 0.87432  & 0.17820   & 1.15726 & 2.40732    \\
Nb       & 0.85097  & 0.99247  & 0.24572   & 1.08802 & 2.19244    \\
Mo       & 0.83057  & 0.97345  & 0.10805   & 1.06314 & 2.01708    \\
Tc       & 0.64629  & 1.08072  & 0.09302   & 1.00014 & 1.91300    \\
Ru       & 0.65130  & 1.07465  & 0.04760   & 1.00001 & 1.80799    \\
Rh       & 0.68579  & 1.06923  & 0.06445   & 0.99989 & 1.71702    \\
Pd       & 0.57192  & 0.95218  & 0.04268   & 0.90172 & 1.63401    \\
Ag       & 0.44541  & 0.79565  & -0.04959  & 0.84306 & 1.52479    \\
         &          &          &           &         &            \\
Ba       & -0.04951 & 0.03502  & -0.20497  & 1.07269 & 3.56198    \\
Hf       & 0.29999  & 0.75423  & 0.32239   & 0.88204 & 2.50856    \\
Ta       & 0.70455  & 0.92990  & 0.23577   & 1.12532 & 2.31790    \\
W        & 0.64038  & 0.86882  & 0.09170   & 1.11008 & 2.17888    \\
Re       & 0.60996  & 1.15988  & 0.15348   & 1.14822 & 2.08878    \\
Os       & 0.53044  & 1.06428  & 0.05117   & 1.11453 & 2.01287    \\
Ir       & 0.47125  & 1.01759  & 0.01404   & 1.06585 & 1.91597    \\
Pt       & 0.43374  & 0.94903  & 0.00569   & 1.00933 & 1.83802    \\
Au       & 0.37521  & 0.84519  & -0.02211  & 0.94002 & 1.75060    \\
Hg       & 0.36137  & 0.68747  & -0.10952  & 0.90569 & 1.53337    \\ \hline
Fe\footnotemark[1] 
         & 0.87761  & 0.84369  & 0.02940  & 0.94826 & 1.33156     \\
Fe\footnotemark[2]  
         & 0.84395  & 0.88024  & 0.19670  & 0.93012 & 1.43124     \\
Ni\footnotemark[1]
         & 0.45155  & 0.69040  & -0.04560 & 0.72937 & 1.22004     \\
Ni\footnotemark[2]
         & 0.46394  & 0.70316  & -0.00173 & 0.73568 & 1.24548     \\
Co\footnotemark[1]
         & 0.69846  & 0.68425  & -0.06187 & 0.79695 & 1.26137     \\
Co\footnotemark[2]
         & 0.66026  & 0.70002  & 0.06184  & 0.77917 & 1.33184     \\
\end{tabular}
\end{ruledtabular}
\footnotetext[1]{Ferromagnetic spin up}
\footnotetext[2]{Ferromagnetic spin down}
\end{table}


\begin{figure}[ht]
\begin{minipage}[c]{.24\textwidth}
\centering
\includegraphics[width=1.5in,height=1.7in]{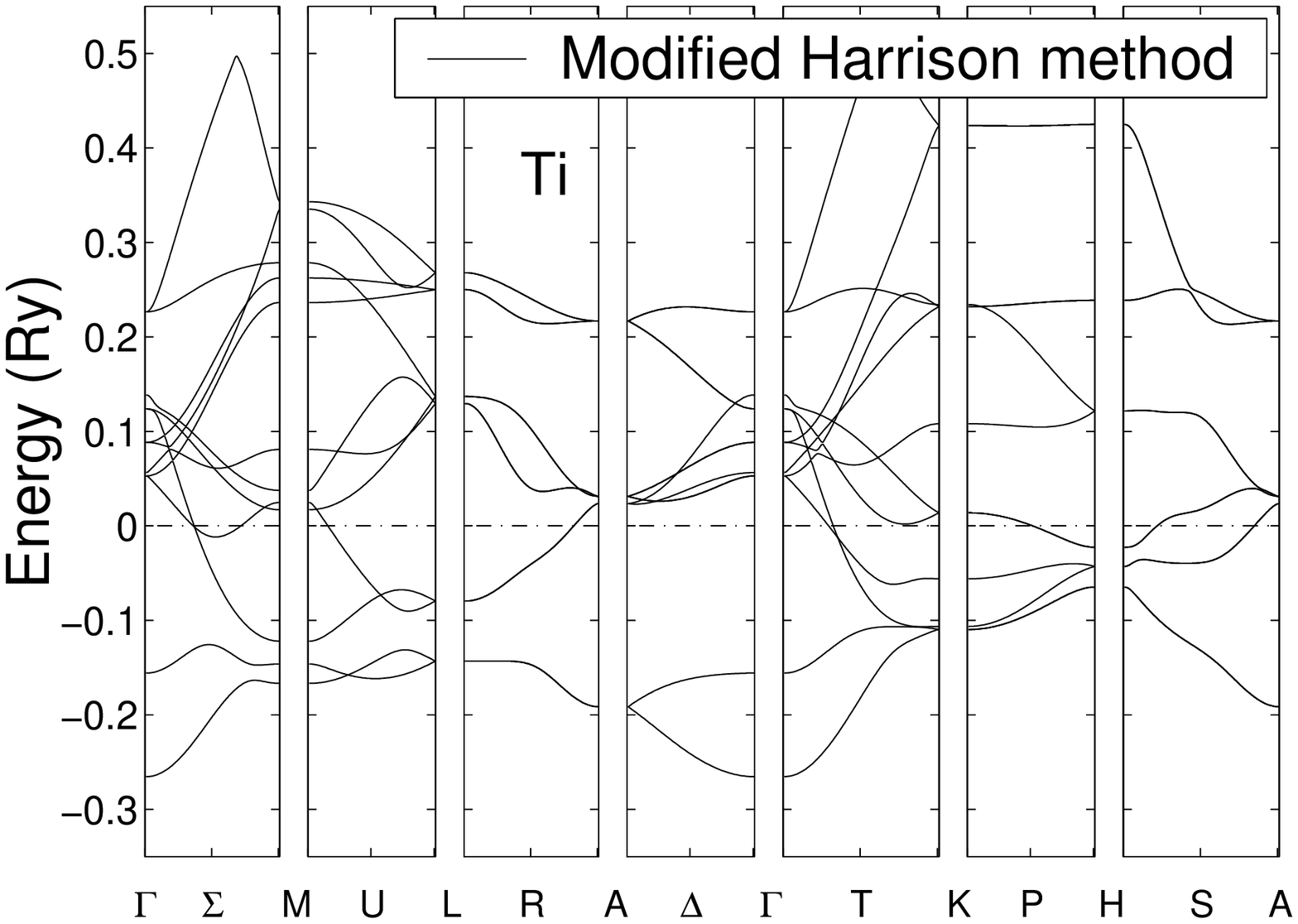}
\end{minipage}%
\begin{minipage}[c]{.24\textwidth}
\centering
\includegraphics[width=1.5in,height=1.7in]{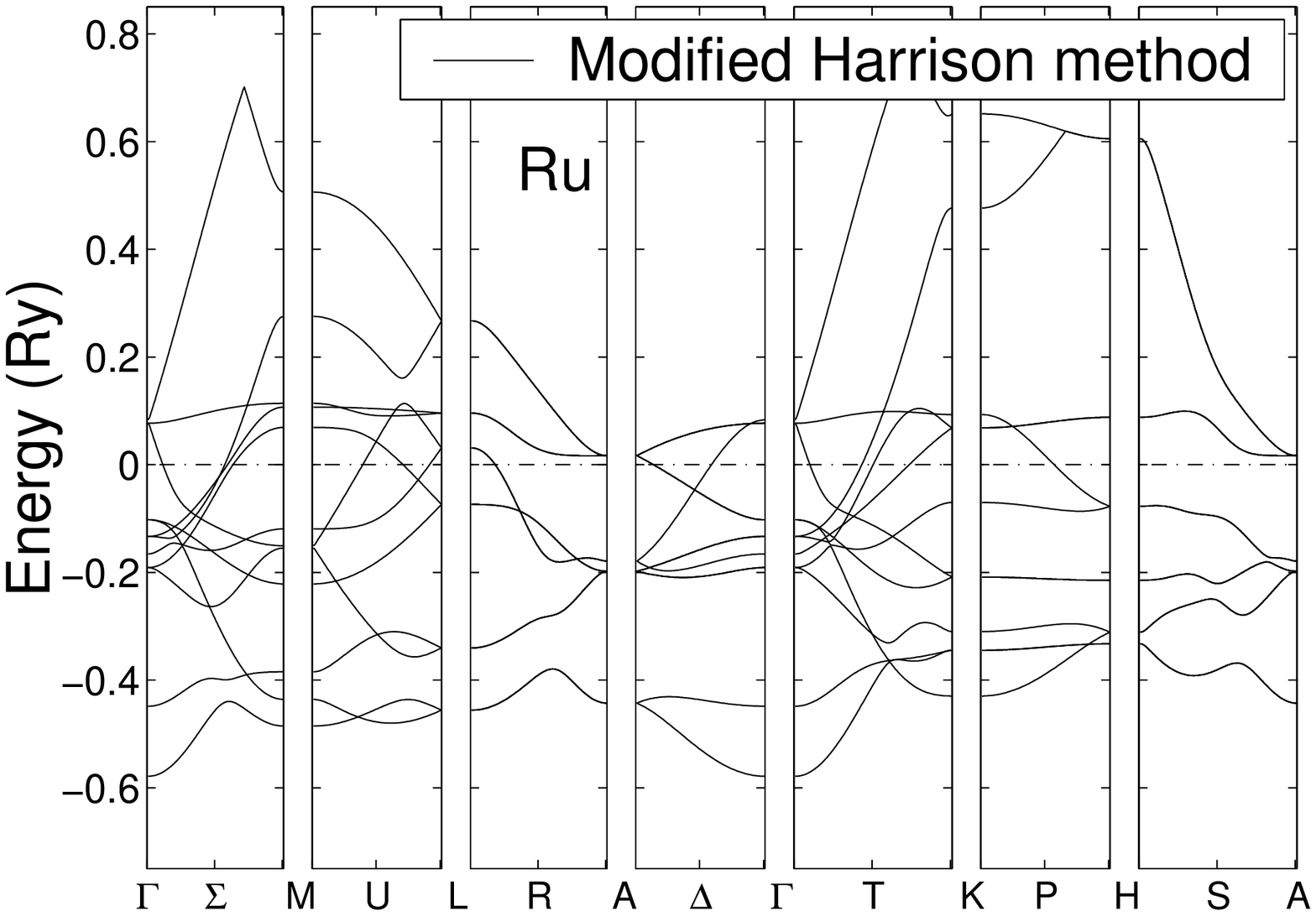}
\end{minipage}%
\caption{Energy bands of hcp Ti and Ru. }\label{band_ti_ru}
\end{figure}

We used the prefactors of Table~\ref{we_sst} with new onsite energies 
\( \gamma_{s} \) and \( r_{d} \) to fit the rest of the transition 
metals, including those with hcp ground states. For the hcp metals,
we fitted energy bands of fcc structures at the
equilibrium lattice, and found that our parameters produce
good transferability, ie. reproduced the hcp energy bands very well 
without fitting them.
The hcp energy bands of Ti and Ru are shown in Fig.~\ref{band_ti_ru}.   
We also fitted energy bands of the ferromagnetic elements Fe, Co and Ni, 
and calculated magnetic moments of the three elements at the experimental
lattice constant. Table~\ref{mag_moment} shows good agreement of 
magnetic moments of Fe, Co and Ni with experimental values.  


\begin{table}[H]
\caption{Magnetic Moments of Fe, Co and Ni.}\label{mag_moment}
\begin{ruledtabular}
\begin{tabular}{|c|c|c|c|} 
Element  & Structure   & TB($\mu_{B}$) & Exp.($\mu_{B}$) \\ \hline
Fe       & bcc         & 2.21     & 2.22       \\
Co       & hcp         & 1.52     & 1.72       \\
Ni       & fcc         & 0.56     & 0.61       \\
\end{tabular}
\end{ruledtabular}
\end{table}


\section{Total Energy}
Next we address the issue of fitting total energy results. In order to 
do this we follow the Naval Research Laboratory tight-binding(NRL-TB) 
methodology \cite{NRL-TB,NRL-review}, which uses parameters which 
are transferable between structures \cite{NRL-TB}. In most TB
approaches as well as in all the so-called ``glue'' potential atomistic
methods one writes the total energy as a sum of a band energy term(
sum of eigenvalues) and a repulsive potential \( G[n(r)] \) that can
be viewed as replacing all the charge density \( n(r) \) dependent
terms appearing in the total energy expression of the density
functional theory. The NRL-TB method has the unique feature that
eliminates G by the following ansatz. We define a quantity \( V_{0} \):
                                                                                
\begin{equation}\label{v0}
V_{0}=\frac{G[n(r)]}{N_{e}}
\end{equation}
                                                                                
where \( N_{e} \) is the number of valence electrons.

We then shift all first-principles eigenvalues \( \epsilon_{i}(k) \)
by the constant \( V_{0} \) and define a shifted eigenvalue:
                                                                                
\begin{equation}\label{shift_e}
\epsilon_{i}^{\prime}(k)=\epsilon_{i}(k)+V_{0}
\end{equation}
                                                                                
The results of this manipulation is that the first-principles
total energy \( E \) is given by the expression:
                                                                                
\begin{equation}\label{new_e}
E=\Sigma \epsilon_{i}^{\prime}(k)
\end{equation}

We note the constant \( V_{0} \) is different for each volume and
structure of the first-principles database. The reader should recognize
that we have shifted each band structure by a constant, retaining
the exact shape of the first-principles bands. It should also be
stressed that all this is done to the first-principles database before
we proceed with the fit that will generate the TB Hamiltonian.
In our trearment of ferromagnetic systems the total energy is equal to
the sum of spin up and spin down shifted eigenvalues. The difference
of these two sums could be viewed as representing the exchange energy. 

We write the onsite energies in a polynomial form:
\begin{equation}
h_{il}(\rho_{i})=\alpha_{l}+\beta_{l}\rho^\frac{2}{3}_{i}+\gamma_{l}\rho^\frac{4}{3}_{i}+\delta_{l}\rho^2_{i}
\end{equation}

where \( l \) is an angular momentum index, and \( \rho_{i} \) is 
an atomic-like density that has the form: 
\begin{equation}
\rho_{i}=\sum_{j\neq i}exp[-\lambda^2R_{ij}]F_{c}(R_{ij})
\end{equation}

where, \( R_{ij} \) is the distance between atom i and j, and \( F_{c}
\) is a smooth cut-off function that was used to limit the range of
parameters \cite{NRL-TB}
\begin{equation}
F_{c} (R)=(1+e^{\frac{R-R_{0}}{R_{l}}})^{-1}
\end{equation}

We take \( R_{0} \) to be in the range of \( 10.0a_{0} 
 \sim  14.0a_{0} \), and \( R_{l}=0.5a_{0} \) ( where \( a_{0} \) 
is Bohr radius), which effectively zeros all interactions for neighbors
more than \( 14.0a_{0} \) apart. Typically, depending on the 
structure and lattice constant, this cut-off function will include 
\( 50 \sim 80 \) neighboring atoms.  

\smallskip

The parameters \( \lambda\),
\(\alpha_{l}\), \( \beta_{l}\), \( \gamma_{l}\)  and  \( \delta_{l} \) 
are determined by fitting total energies following the NRL-TB procedure 
as stated above. The hopping parameters 
were calculated using the modified prefactors of Table~\ref{we_sst}.

\begin{figure}[htbp]
\begin{minipage}[c]{.24\textwidth}
\centering
\includegraphics[width=1.5in,height=1.7in]{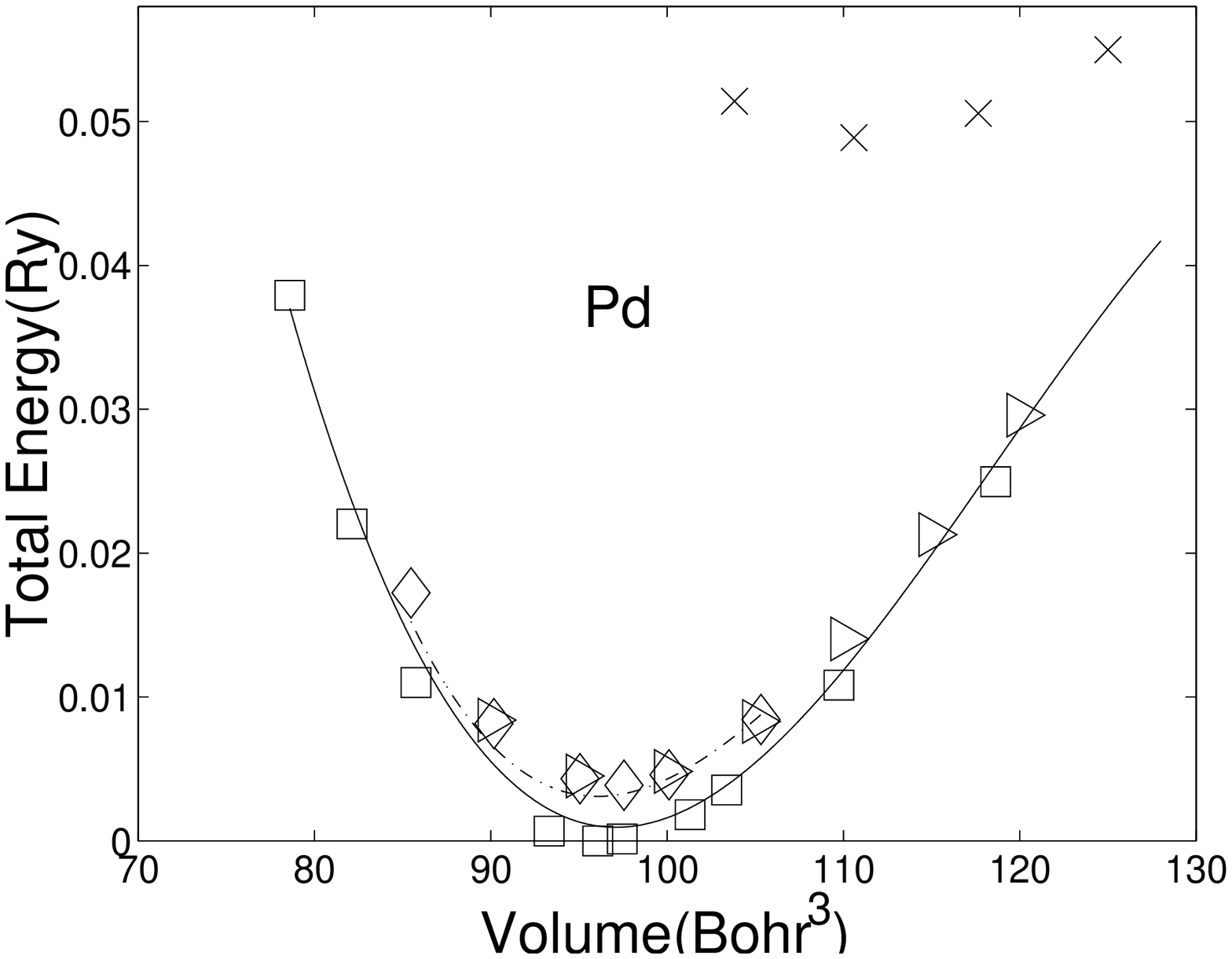}
\end{minipage}%
\begin{minipage}[c]{.24\textwidth}
\centering
\includegraphics[width=1.5in,height=1.9in]{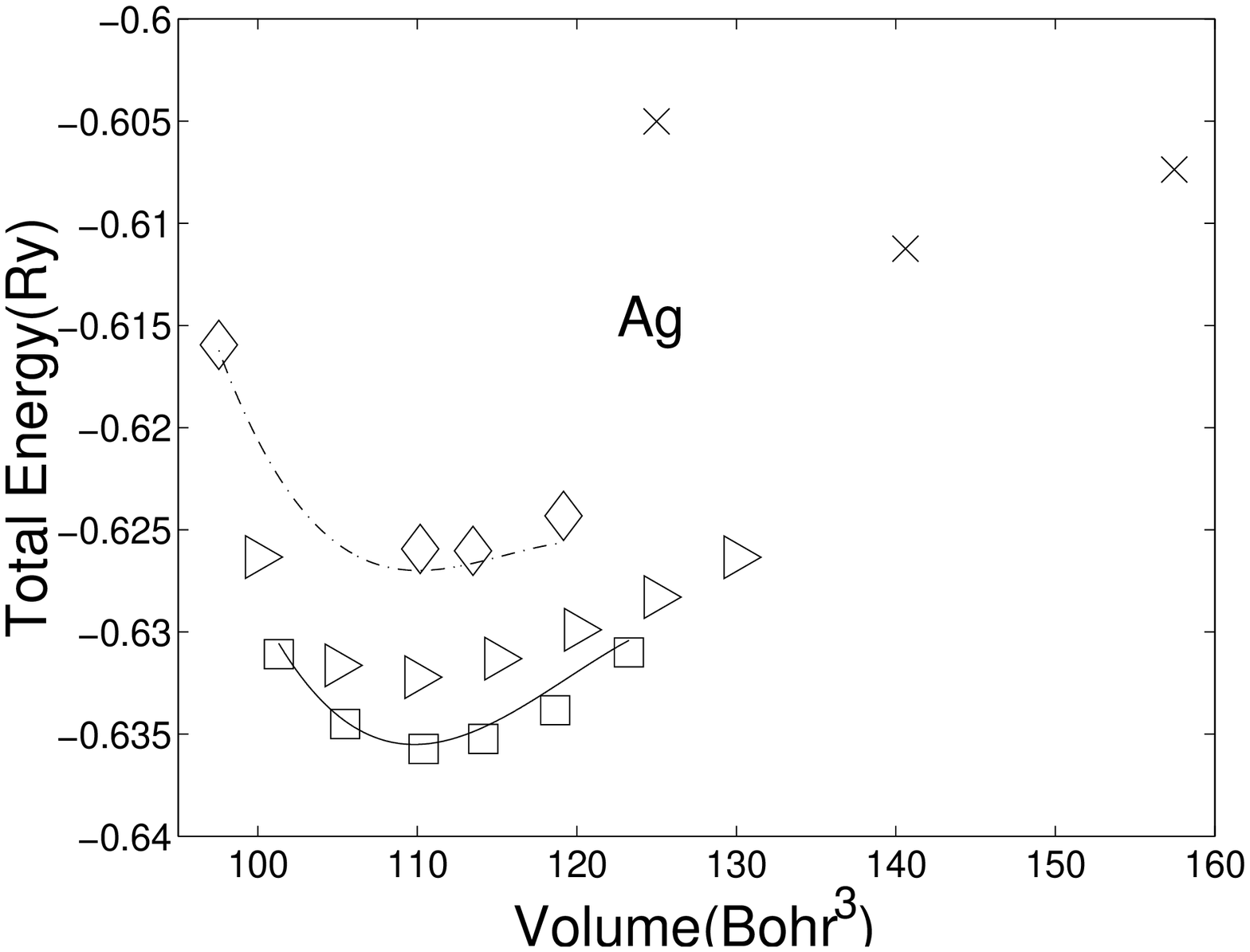}
\end{minipage}%
                                                                                
\begin{minipage}[c]{.24\textwidth}
\centering
\includegraphics[width=1.5in,height=1.7in]{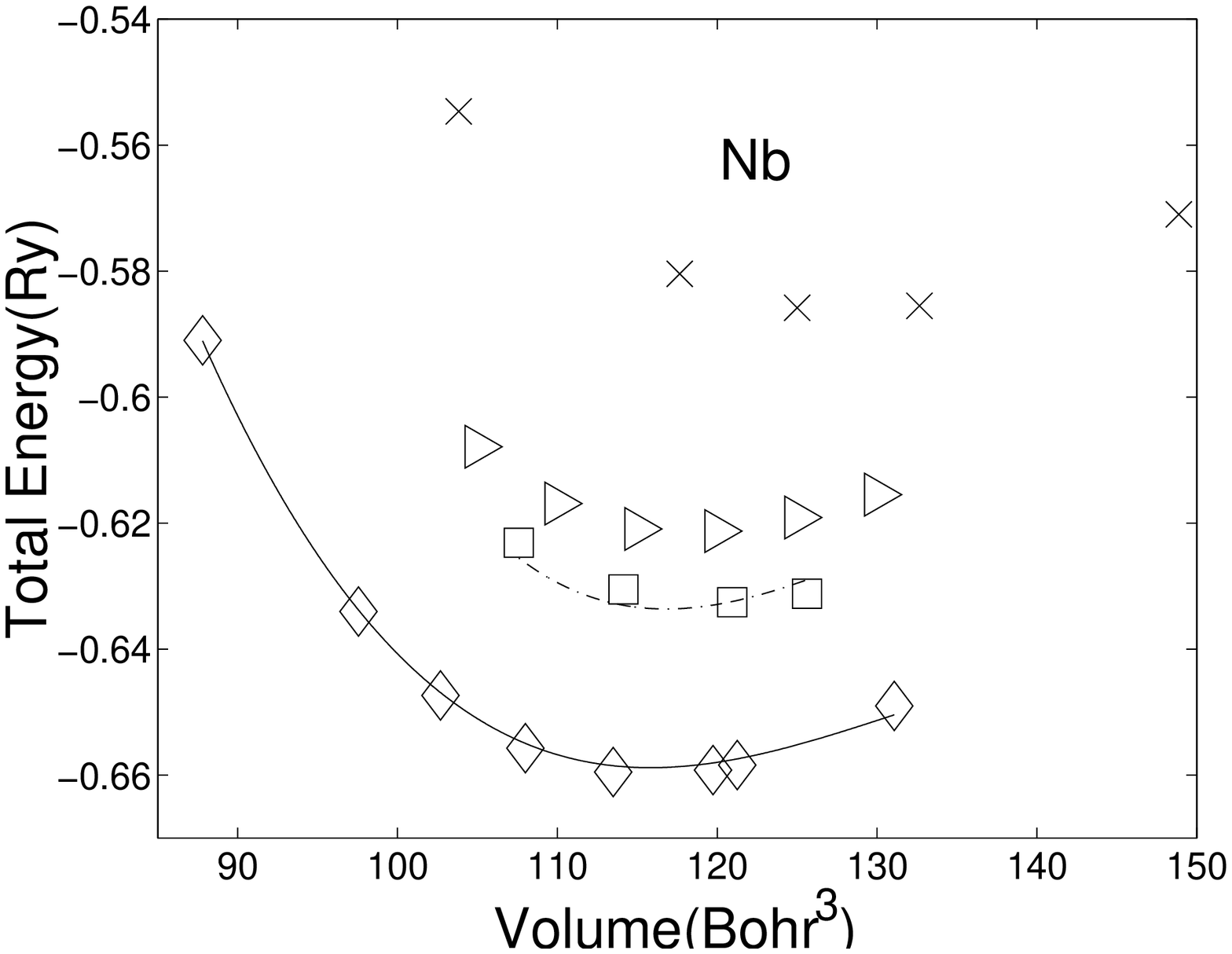}
\end{minipage}%
\begin{minipage}[c]{.24\textwidth}
\centering
\includegraphics[width=1.5in,height=1.7in]{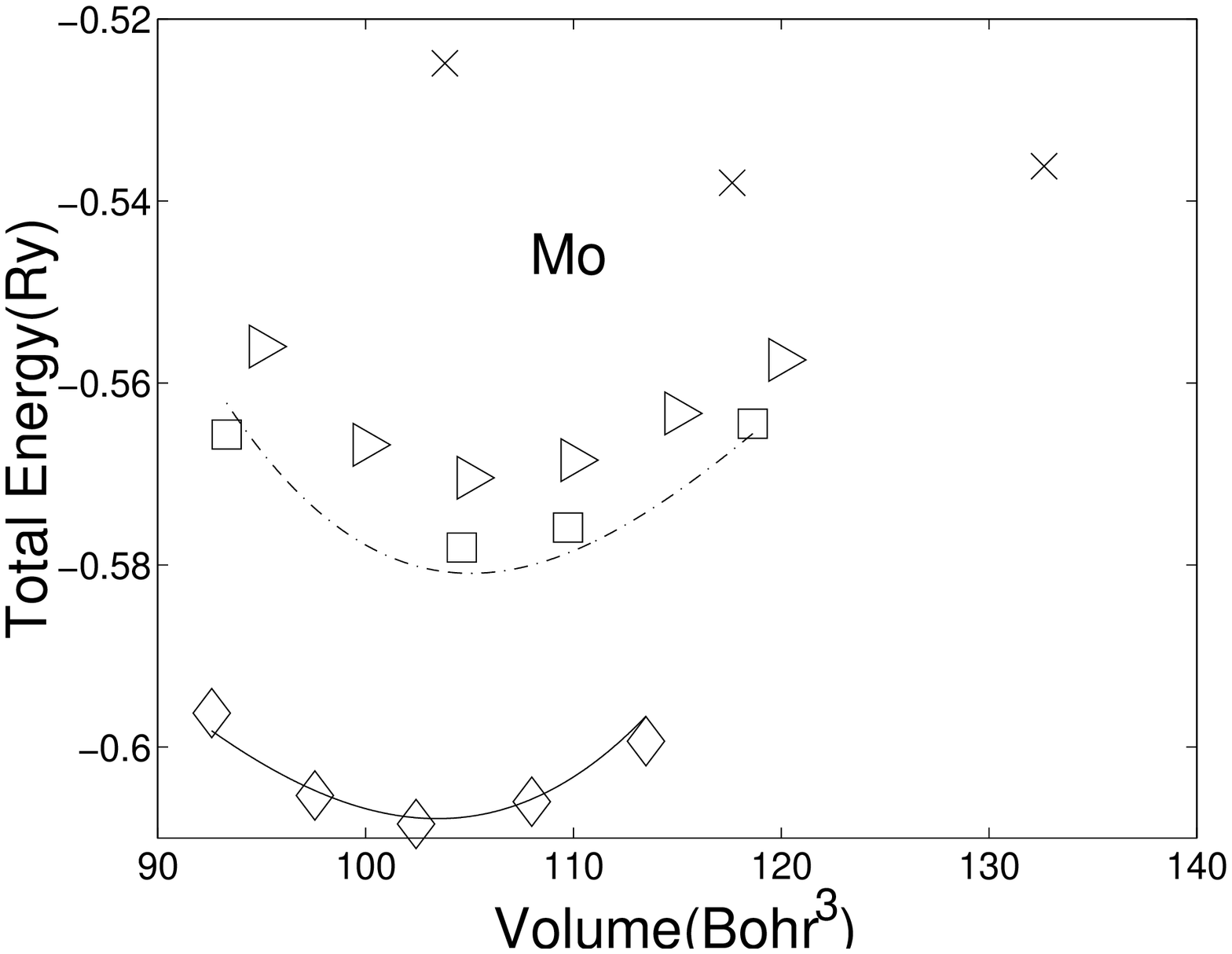}
\end{minipage}
\caption{ Total energies of Ag, Pd, Nb and Mo. The solid line denotes 
bcc APW results and dash-dotted line denotes fcc APW results. The 
diamond, square, triangle and X symbols represent bcc, fcc, hcp and 
sc TB results respectively. }\label{eng_we}
\end{figure}

\begin{figure}[ht]
\begin{minipage}[c]{.24\textwidth}
\centering
\includegraphics[width=1.5in,height=2.0in]{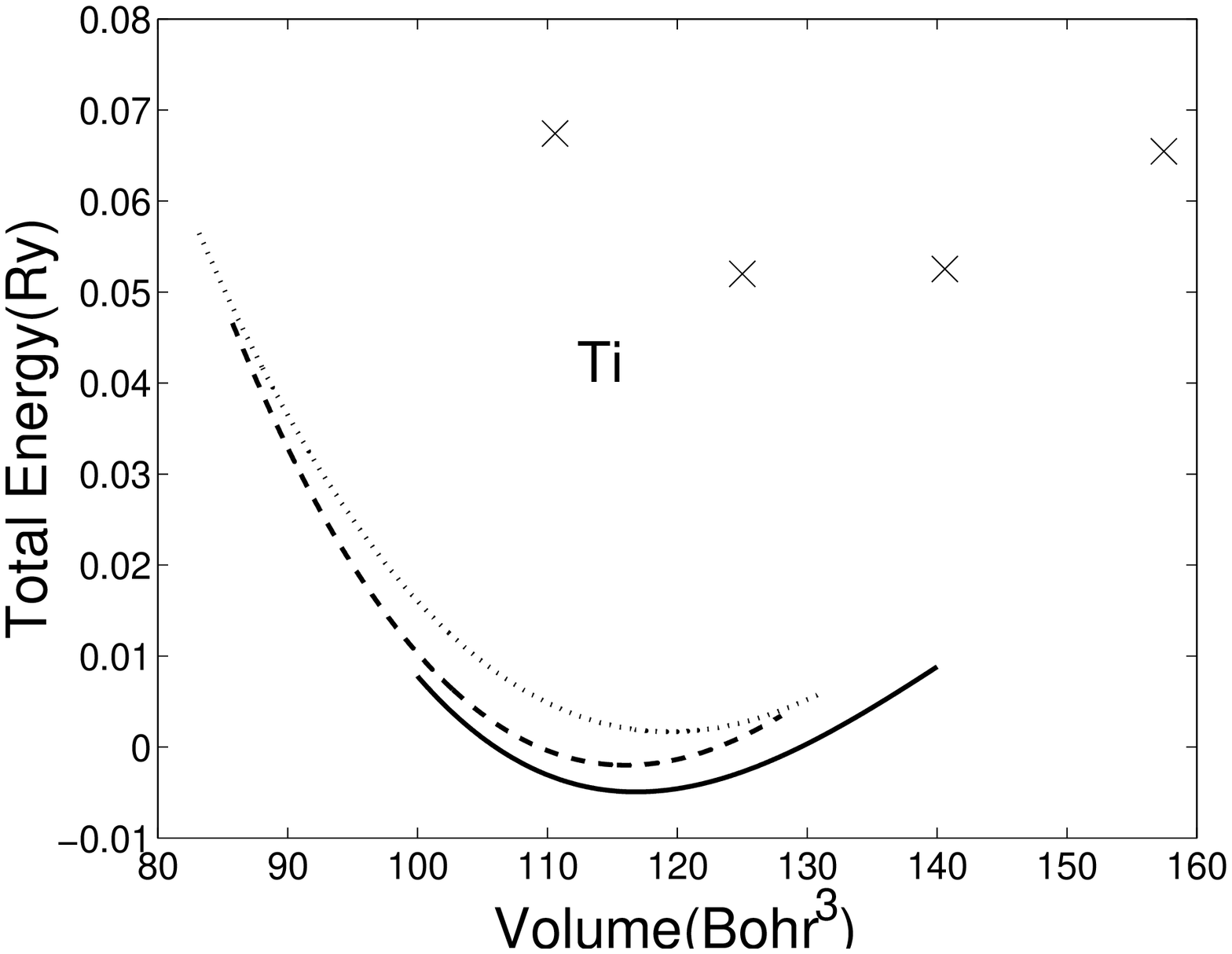}
\end{minipage}%
\begin{minipage}[c]{.24\textwidth}
\centering
\includegraphics[width=1.5in,height=2.1in]{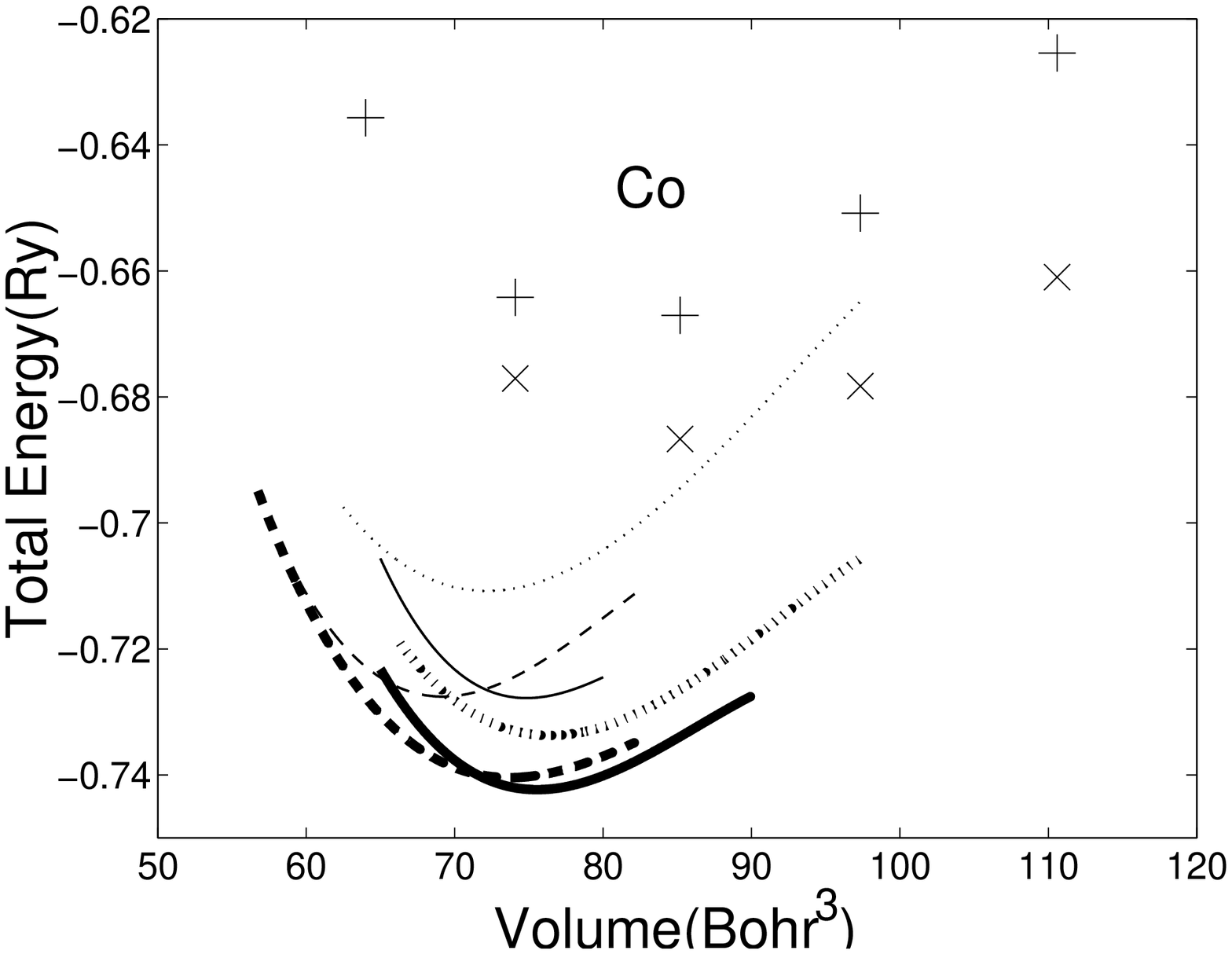}
\end{minipage}
\caption{ TB total energies of Ti and Co. The solid line, dotted line
and dashed line represent hcp, bcc and fcc total energies respectively. 
The symbol + denotes simple cubic total energy. In the right
graph, the thick lines and thin lines represent ferromagnetic and 
paramagnetic Co, and the symbols X and + show ferromagnetic and 
paramagnetic simple cubic respectively. }\label{eng_hcp}
\end{figure}

We fitted total energies of all transition metals to the 
APW results \cite{Sigalas} at several lattice constants of bcc, fcc and 
sc structures. We successfully reproduced 
the ground-state, the order of crystal structures and the 
bulk modulus. Our parameters also place the energies of hcp and sc 
structures, which we did not fit, at  
reasonable values. As an example, we present 
energy-volume relationships for four transition metals in Fig.~\ref{eng_we}, 
and for the hcp metals Ti and Co in Fig.~\ref{eng_hcp}, which again
show the correct ordering of crystal structures. 
We also present in Table~\ref{bulk_we} the equilibrium lattice constants 
and bulk moduli of transition metals.  
 
An inspection of Table~\ref{bulk_we} reveals that our approach matches
very well the LDA lattice constants underestimating the experimental values
by \( 1 \-- 2\% \) for the transition metals and by \( 4 \-- 5\% \) for 
the alkaline earths. The bulk moduli have larger deviation from experiment as 
is usually 
the case in the LDA. For the hcp metals both the lattice parameters and 
bulk moduli are also within the 
LDA predictions except for Tc, Os and Y. Those results could be improved if we 
include the hcp lattice in the fitting database.  


\begin{table}[htbp]
\caption{Equilibrium lattice constants and bulk moduli for the
experimentally \cite{Kittel} observed ground-state structures of the 
elements,
comparing the results of first-principles and tight-binding 
parametrization results.}\label{bulk_we}
\begin{ruledtabular}
\begin{tabular}{|c|c|c|c|c|c|c|c|} 
         & & \multicolumn{3}{c|}{ a(Bohr)} & \multicolumn{3}{c|}{ $B_{0}(Mbar)$} \\ \cline{3-8}
Name  & Structure  & TB   & LDA     & Expt. & TB   & LDA    & Expt. \\ \hline
Ca       & fcc        & 9.98 & 9.96   & 10.55 & 0.21 & 0.13   & 0.15  \\
V        & bcc        & 5.55 & 5.54    & 5.73  & 2.15 & 1.96   & 1.62  \\
Cr       & bcc        & 5.29 & 5.29    & 5.44  & 3.05 & 3.07   & 1.90  \\
Fe\footnotemark[1]    & bcc
                      & 5.38 & 5.38    & 5.43  & 1.76 & 1.76   & 1.68  \\
Ni\footnotemark[1]    & fcc
                      & 6.48 & 6.48    & 6.65  & 2.38 & 2.52   & 1.86  \\
Cu       & fcc        & 6.71 & 6.65    & 6.82  & 2.01 & 1.90   & 1.37  \\
Sr       & fcc        & 10.94 & 10.82  & 11.49 & 0.11 & 0.20   & 0.11  \\
Nb       & bcc        & 6.16 & 6.16    & 6.24  & 1.93 & 1.95   & 1.70  \\
Mo       & bcc        & 5.91 & 5.90    & 5.95  & 2.98 & 2.91   & 2.72  \\
Rh       & fcc        & 7.11 & 7.12    & 7.18  & 3.87 & 3.22   & 2.70  \\
Pd       & fcc        & 7.34 & 7.29    & 7.35  & 1.93 & 1.84   & 1.81  \\
Ag       & fcc        & 7.62 & 7.58    & 7.73  & 1.32 & 1.16   & 1.01  \\
Ba       & bcc        & 9.02 & 9.03    & 9.49  & 0.17 & 0.10   & 0.10  \\
Ta       & bcc        & 6.22 & 6.12    & 6.24  & 2.12 & 2.24   & 2.00  \\
W        & bcc        & 5.99 & 5.94    & 5.97  & 3.63 & 3.33   & 3.23  \\
Ir       & fcc        & 7.30 & 7.29    & 7.26  & 4.14 & 3.86   & 3.55  \\  
Pt       & fcc        & 7.43 & 7.37    & 7.41  & 3.34 & 3.05   & 2.78  \\
Au       & fcc        & 7.77 & 7.67    & 7.71  & 1.87 & 1.70   & 1.73  \\
\hline
         & & \multicolumn{2}{c|}{ a(Bohr)} & \multicolumn{2}{c|}{ c(Bohr)}  & \multicolumn{2}{c|}{ $B_{0}(Mbar)$} \\ \cline{3-8}
Name  & Structure  & TB   & Expt. & TB     & Expt.  & TB     & Expt. \\ \hline 
Sc       & hcp        & 5.98 & 6.25    & 9.55  & 9.96 & 0.34   & 0.44  \\
Ti    & hcp\footnotemark[2]
                      & 5.54 & 5.58    & 8.81  & 8.85 & 1.17   & 1.05  \\
Co\footnotemark[1]    & hcp
                      & 4.74 & 4.74    & 7.70  & 7.69 & 2.35   & 1.91  \\
Y        & hcp        & 6.58 & 6.90    & 10.62  & 10.83 & 0.70   & 0.37  \\
Zr       & hcp        & 5.95 & 6.11    & 9.52  & 9.74 & 0.87   & 0.83  \\
Tc       & hcp        & 5.12 & 5.18    & 8.38  & 8.31 & 5.42   & 2.97  \\
Ru       & hcp        & 5.10 & 5.12    & 7.70  & 8.09 & 3.52   & 3.21  \\
Hf       & hcp        & 6.05 & 6.03    & 9.13  & 9.54 & 1.06   & 1.09  \\
Re       & hcp        & 5.21 & 5.22    & 8.64  & 8.43 & 4.23   & 3.72  \\
Os       & hcp        & 5.28 & 5.17    & 7.61  & 8.16 & 6.98   & 4.18  \\
\end{tabular}
\end{ruledtabular}
\footnotetext[1]{Ferromagnetic}
\footnotetext[2]{hcp lattice fitted}
\end{table}


\section{Conclusion}
To recapitulate, we have accomplished two goals. In the first we 
have reevaluated the ten universal prefactors in Harrison's hopping
parameters and redetermined the \( s, p, d \) onsite energies together
with the parameters \( \gamma_{s} \) and \( r_{d} \). This enables us to 
calculate very accurately the band structure of all the transition, 
alkaline earth and noble metals. For the second goal we have used 
a polynomial form for the 
onsite energies which, with the addition of 15 new parameters, provides 
a total energy capability for our Modified Harrison theory.

Finally, we wish to stress that this work constitutes not only an
efficient computational method but also a valuable addendum to 
Harrison's books.

\smallskip

{\bf Acknowledgment:} We wish to thank Professor Walter A. Harrison and
Drs Michael J. Mehl and Larry L. Boyer for valuable comments.

\end{document}